# The calculation of single-nucleon energies of nuclei by considering two-body effective interaction, $n(k, \rho)$, and a Hartree-Fock *inspired* scheme

H. Mariji[1, 2, †]

[1] *Centro de Física Computacional, Department of Physics, University of Coimbra, P-3004-516 Coimbra, Portugal*
[2] *Centro de Física do Porto, Departmento de Física e Astronomia, Faculdade de Ciências da Universidade do Porto*
*Rua do Campo Alegre,687, 4169-007 Porto, Portugal*

**Abstract**

The nucleon single-particle energies (SPEs) of the selected nuclei; that is, $^{16}$O, $^{40}$Ca, and $^{56}$Ni, are obtained by using the diagonal matrix elements of two-body effective interaction, which generated through the lowest order constrained variational (LOCV) calculations for the symmetric nuclear matter with the $Av_{18}$ phenomenological nucleon-nucleon potential. The SPEs at the major levels of nuclei are calculated by employing a Hartree-Fock *inspired* scheme in the spherical harmonic oscillator basis. In the scheme, the correlation influences are taken into account by imposing the nucleon effective mass factor on the radial wave functions of the major levels. Replacing the density-dependent one-body momentum distribution functions of nucleons, $n(k, \rho)$, with the Heaviside functions, the role of $n(k, \rho)$ on the nucleon SPEs at the major levels of the selected closed shell nuclei, is investigated. The *best* fit of spin-orbit splitting is taken into account when correcting the major levels of the nuclei by using the parameterized Wood-Saxon potential and the $Av_{18}$ density-dependent mean field potential which is constructed by the LOCV method. Considering the point-like protons in the spherical Coulomb potential well, the single-proton energies are corrected. The results show the importance of including $n(k, \rho)$, instead of the Heaviside functions, in the calculation of nucleon SPEs at the different levels, particularly the valence levels, of the closed shell nuclei.

**Keywords**: *Single-particle energy; density-dependent one-body momentum distribution; closed shell nuclei; spin-orbit splitting; Wood -Saxon potential, LOCV, Hartree-Fock inspired scheme.*

## 1. Introduction

The study of shell structure and relevant properties, for example, single-particle energies (SPEs), of closed shell nuclei plays a crucial role in the description of various astrophysical scenarios. For many years, the impact of nuclear shell structure on the astrophysical nucleosynthesis processes has been recognized by observing the so-called *equilibrium*-, *s*-, and *r*-process peaks in the galactic abundance distribution curves [1]. SPEs have frequently been used in a few mass models where the total ground-state energies of finite nuclei are calculated as the sum of a microscopic term, determined by the calculated SPEs, and a macroscopic term, based on the finite range droplet model [2, 3]. Nowadays, the benchmark role of closed shell nuclei properties on a reliable description of astrophysical events becomes more obvious by daily progress in getting the observational data, for example, from the r-process sources such as binary compact stars (white dwarf and neutron star mergers) and explosive supernovae (Ia and II), or from the weak interaction processes in the stellar burning, whether via neutron/electron capture or via neutrino absorption on nucleons/nuclei (see, e.g., [4, 5] and the relevant references therein). Hence, investigating shell structure properties of closed shell nuclei such as SPEs is necessary for the state-of-the-art of the nuclear astrophysics calculations.

---

[†] Email: astrohodjat@fc.up.pt; astrohodjat@gmail.com.

Recently, many works have been presented to describe the shell structure properties of exotic nuclei based on the closed shell nuclei properties. We can point, for instance, to the performing systematic microscopic calculations of medium-mass nuclei [6, 7], the calculations based on microscopic valence-space interactions [8-16], and the calculations by using Self-Consistent Green's Function [17, 18] or In-Medium Similarity Renormalization Group [19, 20]. An important step to study nuclei far from stability is the use of nuclear interactions predicted by chiral effective field theory, which provides consistent two- and three-nucleon forces without any fit to the particular region in the nuclear chart [21, 22]. It is worth mentioning that shell evolutions of various neutron-rich isotopes have been studied by many authors recently, for example, for Si isotopes [23], for Sn isotopes with A=103-110 [24], for Ca isotopes [7], for single-particle-like levels in Sb and Cu isotopes [25]. Lately, the authors of Ref. [26] studied the nuclear states with single-particle and collective characters in the framework of beyond mean field approaches and in the study of semi-magic $^{44}$S. The structure obtained is compatible with results from the state-of-the-art of shell model calculations and experimental data [27-29]. More recently, by applying the newly developed theory of effective nucleon-nucleon interactions on the study of exotic nuclei, the authors of Ref. [30] performed shell-model calculations with several major shells for the exotic neutron-rich Ne, Mg and Si isotopes and predicted the drip lines and showed the effective single-particle energies by exhibiting the shell evolution. Thus, accounting for the shell structure of exotic nuclei, the nuclei far from stability, is essential to both nuclear physics and astrophysics.

Although most of the attempts, mentioned above, lead to results which are compatible with the relevant experimental data, they are not the end of the story in the study of the shell structure of nuclei because of the highly complicated nature of the strong nuclear force. During the past decade, many authors investigated [15, 31, 33-45] how the presence of strong short-range repulsive force in finite nuclei leads to the presence of a high-momentum component in the ground-state wave function, as predicted by scattering experiments. Recently, the high momentum transfer measurements have shown that nucleons in nuclear ground states can form short-range correlated pairs with large relative momentum [45-50]. Furthermore, high-energy electron scattering shows that internucleon short-range repulsive force forms correlated high-momentum neutron-proton pairs and thus, in exotic nuclei, protons are more probable to have a momentum greater than the Fermi momentum in comparing to neutrons [52]. Nevertheless, the effect of the high-momentum component of nuclear force on the properties of nucleons such as nucleon momentum distribution, $n(k)$, and SPE is not well-accommodated in the shell-model theory since this model does not consider the nucleon-nucleon interaction at small distances with precision. For many years, some authors have theoretically predicted the experimental features of $n(k)$, such as the long-tail component in the high-momentum region and the gap at the Fermi surface compared to the fully occupied case of the non-interacting Fermi gas model [52, 53]. However, there are still differences, particularly in the high-momentum region, between experimental $n(k)$ data and those on the basis of theoretical models whether they are microscopic or phenomenological. It is worth mentioning that in the framework of the lowest order constrained variational (LOCV) method [54] and based on the Ristig-Clark formalism [55], the authors of [56] constructed one-body density- and momentum-dependent distribution functions $n(k, \rho)$. Their calculated $n(k, \rho)$ have been in the overall agreement with those of other methods. By applying $n(k, \rho)$ on the calculation of single-nucleon properties, the authors obtained comparable results with other methods [57]. The model-dependent nature of $n(k, \rho)$, particularly in the high-momentum part, leads to disagreements between different methods in the calculation of single-nucleon properties. Hence, the investigation of the influence of high-momentum component on the shell structure properties of nuclei, for example, SPEs, is an open problem in nuclear physics.

In this work, we use the LOCV method to generate $n(k, \rho)$ and two-body effective interaction matrix elements. LOCV is a fully self-consistent technique without any free parameter and with state-dependent correlation functions. The method uses the cluster expansion formalism with an adequate convergence up to two-body interactions. The insignificant effect of higher-order terms, for example, three-body, on the nuclear matter ground-state energy and the convergence of the cluster expansion in the LOCV formalism have been discussed by the authors of [58] and the formalism has been validated. The method has been employed for asymmetric/symmetric nuclear matter (A/SNM) and neutron matter calculations at zero and nonzero temperatures, for different types of neutron stars [59–67], and for finite nuclei calculations from which results have been compatible with experimental data [68-75]. Lately, in order to reduce the effect of the high-momentum component of strong nuclear force on finite nuclei properties, a Fermi momentum cut-off has been imposed on both average effective interactions and density- and channel-dependent ones [76, 77]. More recently, we investigated the effect of the gap of $n(k,\rho)$ at the Fermi surface on the binding energy of the selected medium-mass nuclei [78]. The results of our last work motivate us to investigate the effect of $n(k,\rho)$ on the calculation of the nucleon SPEs of the closed shell nuclei.

There is a problem in the definition and the calculation of single-particle levels (SPLs) for atomic nuclei comprising strongly interacting nucleons. Many years ago, in order to describe experimental measurements, some authors (see, e.g., [79]) attempted to calculate the SPLs of nuclei by presenting a reasonable definition of SPLs. Furthermore, the presence of correlations between interacting nucleons has not been taken into account in the Hartree-Fock (HF) mean field levels [80]. In order to calculate the nucleon SPEs of nuclei, we present a formalized scheme, called here the HF *inspired* scheme, in which the two-body correlations are considered. We include $n(k,\rho)$ in the calculation of density-dependent SPEs and impose a factor of nucleon effective mass, $M^*$ at the Fermi surface, on the radial parts of density functions when extracting SPLs. Thus, we control high-momentum components of two-nucleon interactions by $n(k,\rho)$, rather than using the usual Heaviside functions, and account the presence of correlations in HF mean field by $M^*$. Both $n(k,\rho)$ and $M^*$ are generated by the LOCV calculations for SNM with the $Av_{18}(J_{max}=2)$ phenomenological nucleon-nucleon potential [75]. In order to do this, we organize the paper as follows. In Sec. 2, we formalize a variational calculation of nucleon SPEs at the major levels of nuclei in the spherical harmonic oscillator (SHO) basis. During the calculation, we investigate the role of $n(k,\rho)$ on the calculations at the minimum point of the total energy of a nucleus, denoted by $\gamma_{min}$ ($\gamma$ is the SHO parameter). In Sec. 3, we correct the major levels of nuclei by applying the phenomenological parameterized Wood-Saxon (WS) potential and the $Av_{18}$ mean field potential on the calculation of spin-orbit interaction at $\gamma_{min}$. In Sec. 4, by including the Coulomb interactions we improve the single-proton energies. In Sec. 5, we shall show the effect of $n(k,\rho)$ on the calculation of SPEs in the ground-state of selected closed shell nuclei, i.e., $^{16}$O, $^{40}$Ca and $^{56}$Ni. Finally, in Sec. 6, we discuss the results.

## 2. Calculation of Nucleon SPEs at Major Levels

Although one can obtain the nucleon SPEs of a nucleus by solving the HF self-consistent equations, we calculate them by using a variational method and introducing the HF *inspired* scheme. In our point of view, applying a variational method for the present calculations is a more straightforward approach to show the effect of $n(k,\rho)$ on SPEs. On the other hand, we believe that including a factor of $\sqrt{M^*/M}$ with $M$ as the nucleon mass within the radial parts of SHO functions covers the correlation effects on the mean field calculations. Now, the general form of the Hamiltonian of a nucleus comprising A nucleons is given by:

$$H = \sum_{i=1}^{A} T(i) + \frac{1}{2}\sum_{ij}^{A} V(ij), \qquad (1)$$

where $T(i)$, the single-particle kinetic energy (SPKE) operator for $i$th nucleon, is given by $\hbar^2 k_i^2/2M$ and $V(ij)$ denotes the two-nucleon interaction potential. By denoting the occupied single-particle states by $|c\rangle, |c'\rangle, ...$, the total energy of occupied states (the core energy), $E^c$, in diagonal configuration, is given by

$$E^c = \sum_c^{occ} <c|T|c> + \frac{1}{2}\sum_{cc'}^{occ} <cc'|V|cc'>, \qquad (2)$$

where "$occ$" shows that the summation topped to occupied states. Now, we define the nucleon single-particle interaction energy (SPIE) as the energy which a specific nucleon obtains via the mean field interaction of all other nucleons in the core. By denoting configuration $|c\rangle$ as the state of specific nucleon, SPIE as $U_c = \frac{1}{2}\sum_{c'}^{occ} <cc'|V|cc'>$, and rewriting interaction part of the Hamiltonian, the total energy of core gets following form

$$E^c = \sum_c^{occ} <c|T|c> + \sum_c^{occ} <c|U_c|c>. \qquad (3)$$

Now, we can present an expression to calculate the SPE of a nucleon in the core as follows:

$$\varepsilon_c = <c|T|c> + <c|U_c|c>. \qquad (4)$$

The total energy of a closed shell nucleus with configuration $|c>$ plus one nucleon (one absent-nucleon, so-called, one hole) in the state $|p>$ above (below) the Fermi surface is as follows

$$E^{c \pm p} = E^c \pm <p|T|p> \pm \sum_c^{occ} <cp|V|cp>, \tag{5}$$

By defining $U_p = \sum_c^{occ} <cp|V|cp>$, we can rewrite the total energy as follows:

$$E^{c \pm p} = E^c \pm <p|T|p> \pm <p|U_p|p>. \tag{6}$$

The 1p-1h SPE (near the Fermi surface) reads

$$\varepsilon_{p(h)} = <p|T|p> + <p|U_p|p>. \tag{7}$$

In this work, we set the SHO basis; that is, $|c\rangle = |\alpha; \gamma\rangle$ in which $\alpha = nlm_l$ with $n$, $l$, and $m_l$ as the principle, the orbital angular momentum, and the projection of orbital angular momentum quantum numbers of the nucleon in the core, respectively. The SHO parameter, which plays the role of variation parameter, is given by $\gamma = \sqrt{M\omega/\hbar}$ in which $\hbar\omega$ is the oscillator energy to measure the size of the specific nucleus.

In order to calculate SPEs, our strategy is as follows: first, we obtain the density- and momentum-dependent diagonal matrix elements of the nucleon-nucleon effective interactions, $\mathcal{H}_{eff}(12)$, by using the LOCV calculations for SNM with $Av_{18}(J_{max} = 2)$ potential. Then we employ the elements in order to construct SPIE. Next, while we keep the quantum numbers of one orbit, $\alpha_{c(p)}$, we do summations over the relevant quantum numbers of other orbits in all over occupied states. In order to obtain the expectation values of SPKE and SPIE in configuration space, we use the correlated wave functions by imposing the nucleon effective mass on the radial SHO functions. Finally, we find the minimum value of the total energy of the nucleus, (3), via variation of $\gamma$ and thus the ground-state energy of the nucleus is obtained as we did in our previous works [69-78]. We assign the corresponding $\gamma$ at the minimum point of total energy of the nucleus by $\gamma_{min}$ and calculate the nucleon SPEs of the nucleus at $\gamma_{min}$. During the calculation of SPEs, we shall investigate how the one-body density-dependent momentum distribution functions are important in the evaluation of SPEs by replacing $n(k, \rho)$ with Heaviside functions.

According to (3) and above explanations, the ground-state energy of the core reads

$$E_{g.s}^c(\gamma_{min}) = \sum_{\alpha_c} t(\alpha_c; \gamma_{min}) + \sum_{\alpha_c} u(\alpha_c; \gamma_{min}), \tag{8}$$

where $t(\alpha_c; \gamma_{min}) = <\alpha_c; \gamma_{min}|T|\alpha_c; \gamma_{min}>$ and $u(\alpha_c; \gamma_{min}) = <\alpha_c; \gamma_{min}|U_c|\alpha_c; \gamma_{min}>$ stand for SPKE and SPIE, respectively. Thus, the SPE of nucleon in a specific single-particle state of a closed shell nucleus at $\gamma_{min}$ is given by

$$\varepsilon_{\alpha_c} = t(\alpha_c; \gamma_{min}) + u(\alpha_c; \gamma_{min}), \qquad (9)$$

and the 1p-1h SPE is as follows

$$\varepsilon_{\alpha_{p(h)}} = t(\alpha_p; \gamma_{min}) + u(\alpha_p; \gamma_{min}), \qquad (10)$$

where

$$t(\alpha_p; \gamma_{min}) = <\alpha_p; \gamma_{min}|T|\alpha_p; \gamma_{min}>, \qquad (11a)$$

$$u(\alpha_p; \gamma_{min}) = <\alpha_p; \gamma_{min}|U_p|\alpha_p; \gamma_{min}>. \qquad (11b)$$

In the SHO basis, one can obtain SPIE from (4) and (7) as follows:

$$u_{c(p)}(\gamma) = <\alpha_{c(p)}; \gamma|U_{c(p)}|\alpha_{c(p)}; \gamma> = \left[\frac{1}{2}\right]\sum_{\alpha_2} <\alpha_{c(p)}\alpha_2; \gamma|\mathcal{H}_{eff}(12)|\alpha_{c(p)}\alpha_2; \gamma>, \qquad (12)$$

where $\left[\frac{1}{2}\right]$ means we need a factor of $1/2$ only in the calculation of SPIE for $|\alpha_c>$ (cf. (2)). Regarding the LOCV method, the two-body effective interaction has following form [54]:

$$\mathcal{H}_{eff}(12) = -\frac{\hbar^2}{2M}[F(12), [\nabla^2_{12}, F(12)]] + F(12)V(12)F(12), \qquad (13)$$

where $F(12)$, the density- and channel-dependent two-body correlation functions, are operator type like $Av_{18}$ [73]. The operator $V(12)$ for the $Av_{18}$ potential has following form [81]:

$$V(12) = \sum_\lambda v^\lambda(12) O^\lambda(12), \qquad (14)$$

where the forms of eighteen components of the two-body operator $O^{\lambda=1-18}(12)$ are as follow:

$$O^{\lambda=1-18}(12) = 1, (\sigma_1 \cdot \sigma_2), (\tau_1 \cdot \tau_2), (\sigma_1 \cdot \sigma_2)(\tau_1 \cdot \tau_2), (S_{12}), (\tau_1 \cdot \tau_2)S_{12}, L \cdot S, (\tau_1 \cdot \tau_2)L \cdot S, L^2,$$
$$(\sigma_1 \cdot \sigma_2)L^2, (\sigma_1 \cdot \sigma_2)(\tau_1 \cdot \tau_2)L^2, (L \cdot S)^2, (\tau_1 \cdot \tau_2)(L \cdot S)^2, (T_{12}), (S_{12})(T_{12}), (\tau_{z_1} \cdot \tau_{z_2})$$

$$(15)$$

In Eq. (14), all components $v^\lambda$ are central functions of distance between two interacting nucleons without any dependency on quantum numbers.

Now, the matrix elements of two-body interactions are obtained by sandwiching $\mathcal{H}_{eff}(12)$: (a) between $\sum |\vec{k}_1\vec{k}_2, \eta_1\eta_2><\vec{k}_1\vec{k}_2, \eta_1\eta_2|$ and $\sum |\vec{k}'_1\vec{k}'_2, \eta'_1\eta'_2><\vec{k}'_1\vec{k}'_2, \eta'_1\eta'_2|$, two complete basis sets of plane waves together with spin and isospin parts; (b) between $\int d^3r|\vec{r}\vec{R}><\vec{r}\vec{R}|$ and $\int d^3r'|\vec{r}'\vec{R}'><\vec{r}'\vec{R}'|$, two orthogonal bases in the configuration space, and (c) between $\sum_\lambda |\lambda><\lambda|$ and $\sum_{\lambda'} |\lambda'><\lambda'|$, two complete sets in which $\lambda$ denotes $L$, $S$, $T$, $M_T$, and $J$

where $L, S, T$, and $M_T$ are the relative orbital momentum, the total spin, isospin, and isospin projection of two interacting nucleons, respectively, and $J$ comes from $\vec{J} = \vec{L} + \vec{S}$. Thus, a straightforward calculation, by taking just diagonal matrix elements and considering orthogonality conditions, (12) is simplified by

$$u_{\alpha_{c(p)}}(\rho,\gamma) = \left[\frac{1}{2}\right]\frac{\Omega(\gamma)}{(2\pi)^3}\int d\vec{k}_1 n(k_1,\rho)\left|\phi_{\alpha_{c(p)}}(\vec{k}_1,\gamma^{-1})\right|^2 \left[U_{\alpha_{c(p)}}(k_1,\rho,\gamma)\right], \quad (16)$$

where

$$U_{\alpha_{c(p)}}(k_1,\rho,\gamma) = \frac{1}{(2\pi)^2}\sum_{\alpha_2, m_{\tau_2}}\sum_{\lambda}\left|C^{TM_T}_{m_{\tau_1}m_{\tau_2}}\right|^2 (2J+1)[1-(-1)^{L+S+T}]/2$$

$$\times \left[\int d\vec{k}_2 n(k_2,\rho)|\phi_{\alpha_2}(\vec{k}_2,\gamma^{-1})|^2 \int \mathcal{H}^{\lambda}_{eff}(\mathbb{r},\rho)|j_L(\mathbb{r}\mathbb{k})|^2\mathbb{r}^2 d\mathbb{r}\right]. \quad (17)$$

In (17), $\left|C^{TM_T}_{m_{\tau_1}m_{\tau_2}}\right|$ are the iso-spin Clebsch-Gordon coefficients and $j_L(\mathbb{r}\mathbb{k})$ are the well-known spherical Bessel functions in which the relative momentum and distance of two interacting nucleons are obtained by the familiar relations $\mathbb{k} = |\vec{k}_1 - \vec{k}_2|/2$ and $\mathbb{r} = |\vec{r}_1 - \vec{r}_2|$, respectively. In the above equations, we converted the summations over $\vec{k}_1$ and $\vec{k}_2$ to integrals by the well-known relation as follows:

$$\frac{1}{\Omega}\sum_{\vec{k}} \rightarrow \frac{1}{(2\pi)^3}\int n(k) d\vec{k}, \quad (18)$$

where $n(k)$ are the one-body momentum distribution functions. In this work, $n(k)$, generated by using the LOCV calculations for SNM with $Av_{18}(J_{max} = 2)$ potential based on the Ristig-Clark formalism [55], are density-dependent [78]. We suppose that the volume of box, $\Omega$, is obtained by $\Omega(\gamma) = 4\pi[R_{rms}(\gamma)]^3/3$ where $R_{rms}(\gamma)$ calls

$$R^2_{rms}(\gamma) = \int \rho(R,\gamma) R^4 dR. \quad (19)$$

In (16) and (17), we can replace $\phi_\alpha(\vec{k},\gamma^{-1})$ by $(2l+1)|R_{nl}(k,\gamma^{-1})|^2/4\pi$, since SPKE and SPIE depend only on the magnitude of the relevant single-particle momentum and also the interest of this work is to consider the closed shell nuclei. It should be noted that $k$'s are topped to the Fermi momentum, $K_F(\rho) = (6\pi^2\rho/\nu)^{1/3}$ with $\nu$ as degeneracy parameter of nucleons. Now, by use of (9), (10), (16), and (17), and recalling $<O> = Tr(\rho O)$, we find SPE as follows:

$$\varepsilon_{nl_{c(p)}}(\gamma) = \int dR R^2 |\tilde{R}_{nl}(R,\gamma)|^2 \left[t_{nl_{c(p)}}(\rho,\gamma) + u_{nl_{c(p)}}(\rho,\gamma)\right], \quad (20)$$

where we replaced $\alpha$ with $nl$. In the above equation, $\tilde{R}_{nl}(R,\gamma)$ reads

$$\tilde{R}_{nl}(R,\gamma) = C_{nl}\sqrt{M^*(R)/M}\, R_{nl}(R,\gamma), \quad (21)$$

where $C_{nl}$ is a normalization coefficient for each natural orbit and $M^*(R)$, the effective mass of nucleon, calculated at the Fermi surface by the LOCV method with considering $Av_{18}(J_{max} = 2)$ potential [78]. In this work, we calculate $M^*$ at a fixed energy, so-called *k-mass* definition, which is a measure of the *non-locality* of single-particle potential, due to *non-locality* in space. This definition is clearly different from the *E-mass* definition which shows the non-locality of single-particle potential, due to *non-locality* in time. By imposing this factor we mean the shape of density distribution function is altered by correlations in the HF *inspired* scheme. After varying $\gamma$ and finding the minimum total energy of nucleus, we obtain the SPE of a specific single-particle state at $\gamma_{min}$ ($\varepsilon_{nl_{c(p)}}(\gamma_{min})$). Table 1 shows $\varepsilon_{nl}(\gamma_{min})$ for the single-nucleon major levels of $^{16}$O, $^{40}$Ca, and $^{56}$Ni. In order to clarify the importance of including $n(k,\rho)$ in calculations, the values of $\varepsilon_{nl}$ are considered in two different cases: *(i)* including the constructed $n(k,\rho)$ *(ii)* including the Heaviside function. In Table 1, the magnitudes of SPEs at the major levels, especially near the valence levels, indicate the importance of including $n(k,\rho)$, as discussed in Sec. 6.

**Table 1** The single-nucleon major levels of energies (in MeV) of the sample nuclei, at $\gamma_{min}$ (in $fm^{-1}$) by including two distribution functions: *(i)* $n(k,\rho)$ *(ii)* Heaviside (see text).

| Nuclei | *Distribution* | $\gamma_{min}$ | $\varepsilon_{0s}$ | $\varepsilon_{0p}$ | $\varepsilon_{0d}$ | $\varepsilon_{1s}$ | $\varepsilon_{0f}$ | $\varepsilon_{1p}$ |
|---|---|---|---|---|---|---|---|---|
| $^{16}$O | *(i)* | 0.310 | -23.003 | -11.100 | -5.404 | -14.203 | - | - |
| | *(ii)* | 0.330 | -26.483 | -35.336 | -34.465 | -21.434 | - | - |
| $^{40}$Ca | *(i)* | 0.330 | -22.273 | -11.939 | -7.299 | -15.946 | +0.674 | -1.127 |
| | *(ii)* | 0.360 | -19.140 | -30.205 | -36.423 | -23.177 | -29.531 | -18.033 |
| $^{56}$Ni | *(i)* | 0.340 | -19.094 | -13.321 | -7.644 | -13.661 | -0.581 | -4.993 |
| | *(ii)* | 0.360 | -17.219 | -31.566 | -36.567 | -19.694 | -29.535 | -20.029 |

### 3. Spin-Orbit Interaction

In order to complete the calculations of SPLs and the evaluation of the $n(k,\rho)$ role on them, we include the spin-orbit effect in our calculations. There are many calculations of spin-orbit corrections, from the old calculations to new relativistic ones [82-87]. Considering spin-orbit interaction, the spin orientation degeneracy is removed in the major SPLs and they will be split.

Indeed, the SHO base kets, $|\alpha, \gamma>$, are not appropriate in the presence of spin-orbit interaction. Including total angular momentum quantum number, $j$, the appropriate base kets are $|n(ls)j, m_j, \gamma>$, where $m_j = m_l + m_s$ and $m_s$ is the spin projection. Thus, SPEs at $\gamma_{min}$, corresponding $\gamma$ to the minimum point of total energy of the nucleus, are given by:

$$\varepsilon_{nlj}(\gamma_{min}) = \varepsilon_{nl}(\gamma_{min}) + \Delta\varepsilon_{nlj}(\gamma_{min}), \tag{22}$$

where

$$\Delta\varepsilon_{nlj}(\gamma_{min}) = <n(ls)j, m_j, \gamma_{min}|\hat{V}_{ls}|n(ls)j, m_j, \gamma_{min}>. \tag{23}$$

Sandwiching the spin-orbit operator, $\hat{V}_{ls} \sim \vec{l}.\vec{s}$, between two complete orthogonal sets of position base kets, imposing effective mass factor on the radial SHO functions, and performing straightforward mathematics, the spin-orbit correction is written as:

$$\Delta\varepsilon_{nlj}(\gamma_{min}) = v_{SO} \int [df(R)/dR]|\tilde{R}_{nl}(R, \gamma_{min})|^2 R dR \times \begin{cases} -(l+1)/2 \,; j = |l-1/2| \\ l/2 \quad\quad ; j = l+1/2 \end{cases}. \tag{24}$$

In (24), the forms of constant factor $v_{SO}$ and the WS type function $f(R)$ are as follow [82]:

$$v_{SO} = \lambda V_0 (\hbar/\mu c)^2/2, \tag{25a}$$

$$f(R) = \{1 + \exp[(R - R_0)/a]\}^{-1} \tag{25b}$$

where $\lambda$ is a fit parameter, $V_0$ presents the total strength, $\mu$ is the reduced mass of the single particle and residual nucleus mass, $R_0$ is given by $R_0 = r_0 A^{1/3}$, and $a$ is the skin depth which shows diffuseness of the surface. In order to achieve the *best* agreement with the experimental spin-orbit splitting of nuclei the parameters take following values at $\gamma_{min}$:

$$V_{0,best} = 55.00 \; MeV \;;\; r_{0,best} = 1.25 \; fm \;;\; a_{best} = 0.66 \; fm \;;\; \lambda_{best} = 95.00. \tag{26}$$

As another approach, we use our density-dependent single-particle potential, $u_{\alpha_{c(p)}}(\rho, \gamma)$, as $f(R)$. In this way, we fit the spin-orbit correction with the experimental data only by one parameter; that is, $\lambda$, through $v_{SO} = \lambda(\hbar/\mu c)^2/2$. In order to obtain the *best* agreement with the experimental spin-orbit splitting of nuclei, the parameter $\lambda_{best}$ takes value 0.528 at $\gamma_{min}$. Table 2 shows the results of spin-orbit interaction corrections on the major SPLs of $^{16}$O, $^{40}$Ca and $^{56}$Ni, calculated at $\gamma_{min}$, according to the *best*-fit values for the parameters of WS and $u(\rho, \gamma)$ potentials. We put the values of spin-orbit splitting corresponding to the $Av_{18}$ mean field potential, (16), inside the parentheses. In order to characterize the accuracy of the *best*-adjustment data in the description of the calculated spin-orbit splitting, $\delta_{nl}^{cal} = \Delta\varepsilon_{nl,j=l-1/2} - \Delta\varepsilon_{nl,j=l+1/2}$, the average theoretical error of predictions is given by:

$$AD\varepsilon_{nl} = \sqrt{\frac{1}{N}\sum_{i=1}^{N}(\delta_{nl,i}^{cal} - \delta_{nl,i}^{exp})^2}, \tag{27}$$

where $\delta_{nl,i}^{exp}$ is the empirical spin-orbit splitting of each nucleus. Regarding the lack of empirical spin-orbit splitting data for some nuclei, as seen in Table 2, the value of $N$ is 14 in the above relation.

**Table 2** Spin-orbit corrections (in MeV) on the single-neutron and -proton levels of $^{16}$O, $^{40}$Ca, and $^{56}$Ni at the corresponding $\gamma_{min}$ (in $fm^{-1}$). The corrections have been obtained by employing WS and $u(\rho,\gamma)$ potentials, which the relevant values of latter potential have been put inside parentheses. The average error has been calculated by considering the empirical spin-orbit splitting [31, 32].

| Nuclei | $\gamma_{min}$ | $\delta\varepsilon_{0p}^{cal}$ | $\delta\varepsilon_{0d}^{cal}$ | $\delta\varepsilon_{0f}^{cal}$ | $\delta\varepsilon_{1p}^{cal}$ | $\delta\varepsilon_{0p}^{exp}$ | $\delta\varepsilon_{0d}^{exp}$ | $\delta\varepsilon_{0f}^{exp}$ | $\delta\varepsilon_{1p}^{exp}$ | $AD\varepsilon_{nl}$ |
|---|---|---|---|---|---|---|---|---|---|---|
| $^{16}$O-n | 0.31 | 6.05 (2.69) | 4.89 (4.61) | --- | --- | 6.18 | 5.08 | --- | --- | |
| $^{16}$O-p | 0.31 | 6.07 (2.70) | 4.90 (4.63) | --- | --- | 6.32 | 5.04 | --- | --- | |
| $^{40}$Ca-n | 0.33 | 5.10 (2.55) | 7.22 (4.40) | 5.92 (6.24) | 2.32 (2.49) | --- | 5.63 | 5.71 | 2.00 | 0.37 (0.75) |
| $^{40}$Ca-p | 0.33 | 5.11 (2.56) | 7.24 (4.42) | 5.94 (6.27) | 2.32 (2.50) | --- | 5.40 | 5.69 | 1.75 | |
| $^{56}$Ni-n | 0.36 | 3.38 (2.56) | 6.42 (4.42) | 6.50 (6.30) | 1.26 (2.54) | --- | --- | 7.17 | 1.11 | |
| $^{56}$Ni-p | 0.36 | 3.39 (2.56) | 6.44 (4.44) | 6.52 (7.45) | 1.27 (2.55) | --- | --- | 7.45 | 1.11 | |

## 4. Coulomb Interaction

To impose the electromagnetic effects on the single-proton levels, one may suppose that each proton lies in the repulsive Coulomb potential well, made by the other protons. In fact, the potential is determined by a certain nuclear charge distribution, $\rho_C(R)$, as follows:

$$v_C(R) = \begin{cases} (e/R)\int_0^R \rho_C(R')R'^2 dR' & ; \quad R' \leq R \\ e\int_R^\infty \rho_C(R')R' dR' & ; \quad R' > R \end{cases}, \tag{28}$$

where $e = 1.6 \times 10^{-19}\ C$. In this work, the protons are assumed to be point-like particles and to have the uniform spherical charge distribution with the total value $Ze$. Then the Coulomb potential is given by:

$$v_C(R) = \begin{cases} Ze^2[3 - (R/R_C)^2]/2R_C & ; \quad R \leq R_C \\ Ze^2/R & ; \quad R > R_C \end{cases}. \tag{29}$$

In the above equation, $R_C$, the spherical charge distribution radius, is calculated by $R_C = \sqrt{5/3} R_{rms}(\gamma_{min})$.

**Table 3** The Coulomb interaction corrections (in MeV) on the single-proton levels of $^{16}$O, $^{40}$Ca, and $^{56}$Ni at the corresponding $\gamma_{min}$ (in $fm^{-1}$) by including two functions: *(i)* $n(k,\rho)$ *(ii)* Heaviside.

| Nuclei | $\gamma_{min}$ | $\varepsilon_{C,0s}$ | $\varepsilon_{C,0p}$ | $\varepsilon_{C,0d}$ | $\varepsilon_{C,1s}$ | $\varepsilon_{C,0f}$ | $\varepsilon_{C,1p}$ |
|---|---|---|---|---|---|---|---|
| $^{16}$O (i)  | 0.310 | 0.338 | 0.714 | 0.737 | 0.166 | --- | --- |
| $^{16}$O (ii) | 0.330 | 0.365 | 0.794 | 0.920 | 0.171 | --- | --- |
| $^{40}$Ca (i) | 0.330 | 0.367 | 0.785 | 0.920 | 0.175 | 0.741 | 0.547 |
| $^{40}$Ca (i) | 0.360 | 0.402 | 0.885 | 1.135 | 0.198 | 1.121 | 0.524 |
| $^{56}$Ni (i) | 0.34  | 0.387 | 0.807 | 0.988 | 0.186 | 0.864 | 0.533 |
| $^{56}$Ni (ii)| 0.36  | 0.409 | 0.877 | 1.133 | 0.202 | 1.121 | 0.516 |

Comparing to nuclear interactions, there is no significant difference if one supposes the protons are point-like or they have certain radii. In the latter case, it is sufficient to use $R_C^2 = [R_{rms}(\gamma_{min})]^2 + [R_{int}]^2$ in which $R_{int} \approx 0.8\ fm$ [88]. Table 3 shows the results of imposing Coulomb interaction corrections on the single-proton levels, $\varepsilon_{C,nl}$, of the selected nuclei at the corresponding $\gamma_{min}$ by including $n(k,\rho)$ and the Heaviside functions, characterized with *(i)* and *(ii)*, respectively.

## 5. The Effect of $n(k,\rho)$ on SPEs

Including the spin-orbit corrections which have been calculated by $WS$ potential, $\Delta\varepsilon_{nlj}^{WS}$, and Coulomb corrections, $\varepsilon_{C,nl}$, the nucleon SPEs, $\varepsilon_{nl_j}$, read:

$$\varepsilon_{nl_j}(\gamma_{min}) = \varepsilon_{nl}(\gamma_{min}) + \Delta\varepsilon_{nlj}^{WS}(\gamma_{min}) + \varepsilon_{C,nl}(\gamma_{min}). \tag{30}$$

Table 4 shows the calculated SPLs of $^{16}$O, $^{40}$Ca, and $^{56}$Ni at the corresponding $\gamma_{min}$, obtained by the HF *inspired* scheme and the LOCV calculations for SNM with $Av_{18}(J_{max} = 2)$ by including $n(k,\rho)$ and the Heaviside functions, identified by *(i)* and *(ii)*, respectively, besides the experimental data. The letter *p, n,* and the abbreviated word "*exp*" specify the single-proton, single-neutron, and the experimental levels of each nucleus, respectively. As shown in Table 4, including $n(k,\rho)$ leads to a reduction in the magnitude values of all SPEs with respect to the considering the Heaviside functions. This reduction is more significant for the valence levels. We discuss in Sec. 6 about the results.

## 6. Discussion and Conclusion

The momentum and energy distributions of nucleons are affected by the strong short-range correlations of nucleons in finite nuclei. The fact becomes more important when we calculate the nuclei properties, for example, nucleon SPEs, which play a main role in the description of some astrophysical processes (cf. Section 1). Thus, it would be useful if we measure how much $n(k,\rho)$ can influence the prediction of nucleon SPEs of closed shell nuclei. In this work, we have investigated the issue for the selected closed shell nuclei; that is, $^{16}$O, $^{40}$Ca, and $^{56}$Ni, by replacing $n(k,\rho)$, generated by the LOCV calculations for SNM with $Av_{18}(J_{max}=2)$, with

Table 4 The nucleon SPEs of $^{16}$O, $^{40}$Ca, and $^{56}$Ni by considering *(i)* $n(k,\rho)$ and *(ii)* Heaviside functions.

| Nucleus-Scheme | $\varepsilon_{0s_{1/2}}$ | $\varepsilon_{0p_{3/2}}$ | $\varepsilon_{0p_{1/2}}$ | $\varepsilon_{0d_{5/2}}$ | $\varepsilon_{0d_{3/2}}$ | $\varepsilon_{1s_{1/2}}$ | $\varepsilon_{0f_{7/2}}$ | $\varepsilon_{0f_{5/2}}$ | $\varepsilon_{1p_{3/2}}$ | $\varepsilon_{1p_{1/2}}$ |
|---|---|---|---|---|---|---|---|---|---|---|
| $^{16}$O-p (i) | -22.7 | -12.4 | -6.3 | -6.6 | -1.7 | -14.0 | --- | --- | --- | --- |
| $^{16}$O-p (ii) | -26.1 | -36.9 | -29.9 | -36.1 | -29.7 | -21.2 | --- | --- | --- | --- |
| $^{16}$O-n (i) | -23.0 | -13.1 | -7.1 | -7.4 | -2.5 | -14.2 | --- | --- | --- | --- |
| $^{16}$O-n (ii) | -26.5 | -37.7 | -30.7 | -37.0 | -30.7 | -21.4 | --- | --- | --- | --- |
| $^{40}$Ca-p (i) | -21.9 | -12.8 | -7.7 | -9.3 | -2.0 | -15.7 | -1.1 | +4.8 | -1.3 | +0.9 |
| $^{40}$Ca-p (ii) | -18.7 | -31.0 | -25.9 | -38.8 | -30.0 | -23.0 | -32.2 | -23.3 | -18.2 | -16.1 |
| $^{40}$Ca-n (i) | -22.3 | -13.6 | -8.5 | -10.2 | -3.0 | -15.9 | -1.8 | +4.1 | -1.9 | +0.4 |
| $^{40}$Ca-n (ii) | -19.1 | -31.9 | -26.8 | -39.9 | -31.2 | -23.2 | -33.4 | -24.4 | -18.7 | -16.7 |
| $^{56}$Ni-p (i) | -18.7 | -13.6 | -10.2 | -9.2 | -2.8 | -13.5 | -2.5 | +4.0 | -4.9 | -3.6 |
| $^{56}$Ni-p (ii) | -16.8 | -31.7 | -28.6 | -38.2 | -31.2 | -19.5 | -32.9 | -23.6 | -19.9 | -18.8 |
| $^{56}$Ni-n (i) | -19.1 | -14.4 | -11.1 | -10.2 | -3.8 | -13.7 | -3.4 | +3.1 | -5.4 | -4.1 |
| $^{56}$Ni-n (ii) | -17.2 | -32.6 | -29.4 | -39.3 | -32.4 | -19.7 | -33.1 | -24.8 | -20.4 | -19.3 |

Heaviside functions. By constructing a density- and momentum-dependent mean field potential in the framework of LOCV method and by employing the HF *inspired* scheme in which the nucleon effective mass has been imposed on the natural orbits of nuclei, we have calculated the nucleon SPEs at the major levels of the selected nuclei and measured the effect of $n(k,\rho)$ on the calculations. Table 1 illustrates the nucleon SPEs at major levels of the selected nuclei by considering $n(k,\rho)$ and Heaviside functions. We see that there is an absolute increase around 70%, 30%, and 100% in the energy of valence levels of $^{16}$O, $^{40}$Ca, and $^{56}$Ni, respectively, when we replace $n(k,\rho)$ with the Heaviside functions. In order to obtain the correct values of the sub-shell energies, we have included the spin-orbit splitting corrections by using the parameterized WS potential and our density-dependent single-particle potential, $u(\rho,\gamma)$, as the Thomas type mean-field potential. The scheme incorporating this effect has been considered at $\gamma_{min}$, according to the *best* fit to the empirical spin-orbit splitting data and the results have been presented in Table 2. In the case of WS *best*-fit, the calculated average deviation, $AD\varepsilon_{nl}$, from empirical data is surprising with respect to the available data of empirical spin-orbit splitting for both neutron and proton levels of all selected nuclei. Although there is an advantage in the use of the $Av_{18}$ mean-field potential relative to the WS potential because of the existing only one adjustment parameter ($\lambda$), the spin-orbit splitting energies are not compatible with the experimental data for $0p$ levels of the selected nuclei. In order to correct single-proton levels, we assumed that single point-like proton lies in a uniformly charged spherical potential well with a radius determined by the corresponding RMS radius, $R_{rms}(\gamma_{min})$. As shown in Table 3, whereas these corrections do not lead to a significant change in the energy levels of the nuclei, there is a maximum about 20% and 30% decrease for the energy levels $0d$ (all selected nuclei) and $0f$ ($^{40}$Ca), respectively, by replacing the Heaviside functions with $n(k,\rho)$. Table 4 compares the results of two cases of the SPEs calculations by considering two momentum distribution functions: (*i*) $n(k,\rho)$ and (*ii*) Heaviside at the corresponding $\gamma_{min}$. Regarding the magnitude of SPEs, considering $n(k,\rho)$ leads to a significant reduction with respect to the Heaviside case for the valence levels of the selected closed shell nuclei. The values of SPEs depend on the calculation scheme, which encompasses the way the interactions are regularized, and details of the many-body method. However, the results of the employed scheme in the present work show that how the selection of momentum distribution function is important in the SPE calculations. As a result of the above discussions, by taking into account the depletion at the Fermi surface and the tail of the one-body density- dependent momentum distribution in the calculations, through using $n(k,\rho)$, we find a large difference in the nucleon SPEs at the valence levels and other levels of the selected closed shell nuclei.

In conclusion, $n(k,\rho)$ has been employed to calculate the SPEs of closed shell nuclei, i.e., $^{16}$O, $^{40}$Ca and $^{56}$Ni, by using the matrix elements of nucleon-nucleon effective interaction and $n(k,\rho)$ from the LOCV calculations for SNM with $Av_{18}$ and also employing a HF *inspired* scheme to consider the correlation effects through imposing the nucleon effective mass on the density distribution functions of nucleons. The surprising results show that the use of $n(k,\rho)$ has a significant effect on the calculation of SPEs. The results of the present formalism might be improved by including all interaction matrix elements and constructing the diagonalized interaction matrix, by including three-body force effects, as well as the density-dependent Coulomb interactions. Although the calculations have been limited by the LOCV method (to generate the two-body effective interaction matrix elements), the Ristig-Clark formalism (to construct $n(k,\rho)$ functions), and the HF *inspired* scheme (to include correlation effects), the results show how much the inclusion of $n(k,\rho)$ can be important in the prediction of closed shell properties such as single-nucleon energies.

## Competing Interests

## Acknowledgments

The author wishes to thank the Laboratory for Advanced Computing at the University of Coimbra for providing high performance computing resources in order to calculate the two-body effective interactions matrix elements. The author would also like to thank Professor Constança Providência for her hospitality at the Centro de Física Computacional, Departamento de F´ısica, Universidade de Coimbra, Portugal, Professor Orfeu Bertolami for his hospitality at the Departamento de Física e Astronomia, Faculdade de Ciências, Universidade do Porto, Portugal, Professor I. Vidaña for their good discussions, and Dr. S. Chiacchiera for her useful help on the use of the cluster.

**Reference**

[1] E.M. Burbidge, G.R. Burbidge, W.A. Fowler, and F. Hoyle, Rev. Mod. Phys. **29**, (1957) 547.
[2] H. von Groote, E.R. Hilf, and K. Takahashi, At. Data Nucl. Data Tables **17** (1976) 418.
[3]P. Möller, J. R. Nix, W.D. Myers, and W. J. Swiatecki, At. Data Nucl. Data Tables **59** (1995) 185.

[4] H. Grawe, K. Langanke, G. Martínez-Pinedo, Rep. Prog. Phys. **70**, 1525 (2007).
[5] K. Langanke, F.-K. Thielemann, M. Wiescher, *Nuclear Astrophysics and Nuclei far from Stability*, Lect. Notes Phys. **651**, 383–467 (2004).
[6] T. Otsuka, T. Suzuki, J. D. Holt, A. Schwenk, and Y. Akaishi: Phys. Rev. Lett. **105**, 032501 (2010).
[7] J. D. Holt, T. Otsuka, A. Schwenk, and T. Suzuki: J. Phys. G **39**, 085111 (2012).
[8] A.T. Gallant, J.C. Bale, T. Brunner, U. Chowdhury *et al.*: Phys. Rev. Lett. **109**, 032506 (2012).
[9] F. Wienholtz, D. Beck, K. Blaum, Ch. Borgmann *et al.*: Nature **498**, 346 (2013).
[10] J.D. Holt, J. Menéndez, and A. Schwenk, Eur. Phys. J. A **49**, 39 (2013),
[11] C. Caesar, J. Simonis, T. Adachi, Y. Aksyutina *et al.*, Phys. Rev. C **88**, 034313 (2013).
[12] J.D. Holt, J. Menéndez, and A. Schwenk, Phys. Rev. Lett. **110**, 022502 (2013).
[13] J. D. Holt, J. Menéndez, and A. Schwenk: J. Phys. G **40**, 075105 (2013),
[14] J. D. Holt, J. Menéndez, J. Simonis, and A. Schwenk, Phys. Rev. C **90**, 024312 (2014).
[15] G. Hagen, M. Hjorth-Jensen, G. R. Jansen, R. Machleidt, and T. Papenbrock: Phys. Rev. Lett. **109**, 032502 (2012).
[16] R. Roth, S. Binder, K. Vobig, A. Calci *et al.*, Phys. Rev. Lett. **109**, 052501 (2012).
[17] V. Somà, C. Barbieri, and T. Duguet, Phys. Rev. C **87**, 011303 (2013).
[18] V. Somà, A. Cipollone, C. Barbieri, P. Navrátil, and T. Duguet, Phys. Rev. C **89**, 061301(R) (2014).
[19] K. Tsukiyama, S.K. Bogner, and A. Schwenk, Phys. Rev. Lett. **106**, 222502 (2011).
[20] H. Hergert, S.K. Bogner, S. Binder, A. Calci, Phys. Rev. C **87**, 034307 (2013).
[21] E. Epelbaum, H.-W. Hammer and U.-G. Meißner, Rev. Mod. Phys. **81**, 1773 (2009).
[22] H.-W. Hammer, A. Nogga, and A. Schwenk, Rev. Mod. Phys. **85**, 197 (2013).
[23] S.R. Stroberg, A. Gade, J.A. Tostevin, *et al.*, Phys. Rev. C **91**, 041302(R) (2015).
[24] Erdal Dikmen and Oğuz, Öztürk, Yavuz Cengiz, Commun. Theor. Phys. **63**, 222 (2015).
[25] Y. Utsuno, T. Otsuka, Y. Tsunoda, *et al.*, J. Phys. S. Conf. Proc., **6**, 010007 (2015).
Recent Advances in Shell Evolution with Shell-Model Calculations
[26] J.L. Egido, M. Borrajo, and T. Rodríguez, Phys. Rev. Lett. **116**, 05202 (2016).
Collective and Single-Particle Motion in Beyond Mean Field Approaches
[27] D. Santiago-Gonzalez et al., Phys. Rev. C 83, 061305 (2011).
[28] Y. Utsuno et al, Phys. Rev. Lett. **114**, 032501 (2015).
[29] F. Nowacki and A. Poves, Phys. Rev. C **79**, 014310 (2009).
[30] N. Tsunoda, T. Otsuka, N. Shimizu, M. Hjorth-Jensen, K. Takayanagi, and T. Suzuki, *arXiv:*1601.06442 (2016).
[31] L. Coraggio, N. Itaco, A. Covello, A. Gargano, T.T.S. Kuo, *Phys. Rev.* C **68** (2003) 034320.
[32] N.V. Gnezdilov, I.N. Borzov, E.E. Saperstein, S.V. Tolokonnikov, Phys. Rev C **89**, 034304 (2014).
[33] P. Navrátil, J.P. Vary, B.R. Barret, Phys. Rev. Lett. **84**, 5728 (2000).
[34] F. Arias de Saavedra, G. Có, A. Fabrocini, Phys. Rev. C **63**, 064308 (2001).
[35] S.C. Pieper, R.B. Wiringa, J. Carlson, Phys. Rev. C **70**, 054325 (2004).
[36] D.J. Dean, M. Hjorth-Jensen, Phys. Rev. C **69**, 054320 (2004).
[37] E. Caurier, G. Martínez-Pinedo, F. Nowacki, A. Poves, and A. P. Zuker, Rev. Mod. Phys. **75**, 427 (2005).
[38] E. Epelbaum, Prog. Part. Nucl. Phys. **57**, 654 (2006).
[39] G. Hagen, T. Papenbrock, D.J. Dean, M. Hjorth-Jensen, Phys. Rev. Lett. **101**, 092502 (2008).
[40] S. Fujii, R. Okamoto, and K. Suzuki, Phys. Rev. Lett. **103**, 182501 (2009).
[41] T. Otsuka, T. Suzuki, J.D. Holt, A. Schwenk, Y. Akaishi, Phys. Rev. Lett. **105**, 032501 (2010).
[42] H. Hergert, S. Binder, A. Calci, J. Langhammer, R. Roth, Phys. Rev. Lett. **110**, 242501 (2013).
[43] E. Epelbaum, H. Krebs, T.A. Lähde, D. Lee, Ulf-G. Meißner, G. Rupak, Phys. Rev. Lett. **112**, 102501 (2014).
[44] A. Cipollone, C. Barbieri, P. Navrátil, Phys. Rev.C **92**, 014306 (2015).
[45] J. J. Kelly, *Nucleon Knockout by Intermediate Energy Electrons, in Advances in Nuclear Physics*, edited by J.W. Negele and E. Vogt (Springer US, New York, 1996) Vol. 23, pp 75-294.
[46] W.H. Dickhoff, C. Barbieri, Prog. Part. Nucl. Phys. **52**, 377 (2004).

[47] F. Benmokhtar, *et al*., Phys. Rev. Lett. **94**, 082305 (2005).
[48] K. S. Egiyan, *et al*., Phys. Rev. Lett. **96**, 082501 (2006).
[49] E. Piasetzky, M. Sargsian, L. Frankfurt, M. Strikman, J.W. Watson, Phys. Rev. Lett. **97**, 162504 (2006).
[50] R. Shneor, *et al*., Phys. Rev. Lett. **99**, 072501 (2007).
[51] O. Hen, *et al*., Science **346**, 614 (2014).
[52] A.N. Antonov, P.E. Hodgson, I.Zh. Petkov, *Nucleon Momentum and Density Distributions in Nuclei*, (Clarendon Press, Oxford, 1988).
[53] V. Pandharipande, I. Sick, P.K.A. de Witt Huberts, Rev. Mod. Phys. **69**, 981 (1997).
[54] R.F. Bishop, C. Howes, J.M. Irvine, and M. Modarres, J. Phys. G: Nucl. Phys. **4**, 1709 (1978).
[55] M.L. Risting, J.W. Clark, Phys. Rev. B **14**, 2875 (1976).
[56] M. Modarres, A. Rajabi, H.R. Moshfegh, Nucl. Phys. A **808**, 60 (2008).
[57] M. Modarres, A. Rajabi, Nucl. Phys. A, **867**, 1 (2011).
[58] M. Modarres, A. Rajabi, H.R. Moshfegh, Phys. Rev. C **76**, 064311 (2007).
[59] M. Modarres, J.M. Irvine, J. Phys. G: Nucl. Phys. **5**, 511 (1979 b).
[60] H.R. Moshfegh, M. Modarres, Nucl. Phys A **749**, 4, 130c (2005).
[61] M. Modarres, G.H. Bordbar, Phys. Rev. C **58**, 5, 2781 (1998).
[62] M. Modarres, J. Phys. G **23**, 8, 923 (1997).
[63] M. Modarres, H.R. Moshfegh, Prog. Theor. Phys. **112**, 1, 21 (2004).
[64] H.R. Moshfegh, M. Modarres, J. Phys. G **24**, 4, 821 (1998).
[65] S. Goudarzi and H.R. Moshfegh, Phys. Rev. C **92**, 035806 (2015).
[66] G.H. Bordbar, M. Bigdeli, T. Yazdizadeh, Int. J. Mod. Phys. A **21**, 5991 (2006).
[67] G.H. Bordbar, S.M. Zebarjad, R. Zahedinia, Int. J. Theor. Phys. **48**, 1, 61 (2009).
[68] M. Modarres, J. Phys. G: Nucl. Phys. **10**, 251 (1984).
[69] M. Modarres, H.R. Moshfegh, and H. Mariji, Can. J. Phys. **80**, 911 (2002).
[70] M. Modarres and N. Rasekhinejad, Phys. Rev. C **72**, 014301 (2005).
[71] M. Modarres and N. Rasekhinejad, Phys. Rev. C **72**, 064306 (2005).
[72] M. Modarres, H. Mariji, and N. Rasekhinejad, JPCS **312**, 092043 (2011).
[73] M. Modarres, N. Rasekhinejad, and H. Mariji, Int. J. Mod. Phys. E **20**, 679 (2011).
[74] M. Modarres, H. Mariji, and N. Rasekhinejad, Nucl. Phys. A **859**, 16 (2011).
[75] M. Modarres and H. Mariji, Phys. Rev. C **86**, 054324 (2012).
[76] H. Mariji, Eur. Phys. J. A **50**, 56 (2014).
[77] H. Mariji and M. Modarres, Part. Nucl. Lett. **11**, 3, 245 (2014).
[78] H. Mariji, Eur. Phys. J. A **52**, 109 (2016).
[79] C. Mahaux and R. Sartor, *Advances in Nuclear Physics: Single Particle Motion in Nuclei*, edited by J.W. Negele and E. Vogt (Springer US, Plenum Press, New-York, 1991).
[80] P.B. Frois, Modern topics in electron scattering: *Nucleon distributions and the nuclear many-body problem*, edited by P.B. Frois and I. Sick (World Scientific, USA and UK, 1991).
[81] R. B. Wiringa, V. Stoks, R. Schiavilla, *Phys. Rev. C* **51**, 38 (1995).
[82] N. Bohr and Mottelson, *Nuclear Structure: Single Particle Motion,* vol. **1**, 259 (1969).
[83] N. Schwierz, I. Wiedenhöver, and A. Volya, *arXiv:0709.3525v1* (2007).
[84] W. Koepf and P. Ring, *Z. Phys.* A **339**, 81 (1991).
[85] G. Mairle and P. Grabmayr, *Eur. Phys. J.* A **9**, 313 (2000).
[86] H. Grawe, A. Blazhev, M. Górska, R. Grzywacz, H. Mach, and I. Mukha, *Eur. Phys. J.* A **27**, s01, 257 (2006).
[87] S.C. Pieper, V.R. Pandharipande, Phys. Rev. Lett. **70**, 2541 (1993).
[88] I.J. Thompson, F.M. Nunes, *Nuclear Reactions for Astrophysics: Principles, Calculation and Applications of Low-Energy Reactions* (Cambridge University Press, London, 2009).